\documentclass{aa}  
\usepackage{graphicx}
\usepackage{txfonts}
\usepackage{natbib}
\usepackage{rotating}
\bibpunct{(}{)}{;}{a}{}{,} 
\begin{document}
  \title{The baryonic content and Tully-Fisher relation at z$\sim$0.6}


  \author{M. Puech \and F. Hammer \and H. Flores \and R.
    Delgado-Serrano \and M. Rodrigues \and Y. Yang }

   \offprints{mathieu.puech@obspm.fr}

\institute{
GEPI, Observatoire de Paris, CNRS, University Paris Diderot; 5 Place Jules Janssen, 92190 Meudon, France 
}

\date{Received ......; accepted .....}

\abstract{Using the multi-integral-field spectrograph GIRAFFE at VLT,
  we previsouly derived the stellar-mass Tully-Fisher Relation (smTFR)
  at $z \sim 0.6$ for a representative sample of 63 emission-line
  galaxies. We found that the distant relation is systematically
  offset by roughly a factor of two toward lower masses from the local
  relation.}{We extend the study of the evolution of the TFR by
  establishing the first distant baryonic TFR in a CDFS subsample of
  35 galaxies. We also investigate the underlying cause of the large
  scatter observed in these distant relations. }{To derive gas masses
  in distant galaxies, we estimate a gas radius and invert the
  Schmidt-Kennicutt law between star formation rate and gas surface
  densities. We consider the influence of velocity dispersion on the
  scatter of the relation, using the kinematic tracer $S$ suggested by
  Kassin and collaborators.}{We find that gas extends farther out than
  the UV light from young stars, a median of $\sim$30\%. We present
  the first baryonic TFR (bTFR) ever established at intermediate
  redshift and show that, within an uncertainty of $\pm$0.08 dex, the
  zeropoint of the bTFR does not appear to evolve between z$\sim$0.6
  and z=0. On the other hand, we confirm that the difference between
  the local and distant smTFR is significant, even considering random
  and systematic uncertainties, and that accounting for velocity
  dispersion leads to a significant decrease in the scatter of the
  distant relation.}{The absence of evolution in the bTFR over the
  past 6 Gyr implies that no external gas accretion is required for
  distant rotating disks to sustain star formation until z=0 and
  convert most of their gas into stars. Finally, we confirm that the
  larger scatter found in the distant smTFR, and hence in the bTFR, is
  caused entirely by major mergers. This scatter results from a
  transfer of energy from bulk motions in the progenitors, to random
  motions in the remnants, generated by shocks during the merging.
  Shocks occurring during these events naturally explain the large
  extent of ionized gas found out to the UV radius in z$\sim$0.6
  galaxies. All the results presented in this paper support the
  ``spiral rebuilding scenario'' of Hammer and collaborators, i.e.,
  that a large fraction of local spiral disks have been reprocessed
  during major mergers in the past 8 Gyr.}


\maketitle

\section{Introduction}
The stellar-mass Tully-Fisher Relation (smTFR) has received increased
attention over the past decade, both at low (e.g.,
\citealt{bell01,pizagno07,meyer08}) and high redshift
\citep{conselice05,flores06,atkinson07,kassin07,puech08,cresci09}.
However, the smTFR does not allow us to homogeneously characterize all
galaxy morphologies together. \cite{mcgaugh00} showed that dwarf
galaxies, which have a larger gas content that more massive galaxies,
fall downward the smTFR. Considering the baryonic (i.e., stellar plus
gas) content of galaxies, he showed that all galaxies, including
dwarves, define a common TFR (see also \citealt{begum08}), which by
then was dubbed the baryonic TFR (bTFR). Because the bTFR appears to
hold over five decades in mass \citep{mcgaugh00}, it can therefore be
considered as being somehow more ``fundamental'' than the smTFR.

So far, the bTFR has been studied only in the local Universe (see
\citealt{verheijen01,bell01,gurovich04,mcgaugh04,mcgaugh05,geha06,noordermeer07,derijcke07,avila08,meyer08,stark09,trachternach09}).
This was primarily because HI observations of galaxies are practically
limited to z$<$0.3 \citep{lah07}. To estimate the gas content of more
distant galaxies, indirect methods were developed, such as the
inversion of the Schmidt-Kennicutt (SK) law \citep{kennicutt89}, which
relates the gas surface and star formation rate (SFR) densities (e.g.,
\citealt{erb06,mannucci09}). Another difficulty in establishing a
reliable bTFR (or even a smTFR) at high z is that it is not always
very clear whether all distant samples are truly representative of the
luminosity or mass function of galaxies at those epochs. At
intermediate redshifts (i.e., z$<$1), \cite{flores06} and
\cite{yang07} assembled a representative sample of 63 z$\sim$0.6
emission-line galaxies observed by 3D spectroscopy for a project
called IMAGES. The kinematics of the entire galaxy surface is probed
by 3D spectroscopy, which allows us to classify them as a function of
their relaxation state. Rotating Disks (RDs) are galaxies that exhibit
regular rotation well-aligned along the morphological axis. Perturbed
rotators (PRs) have large-scale rotation with a local perturbation in
the velocity dispersion map that cannot be accounted for by rotation,
while galaxies with complex kinematics (CKs) do not exhibit
large-scale rotation, or have a strong misalignment between the
dynamical and morphological axes \citep{yang07}. Using this new
kinematical classification, \cite{flores06} showed that the larger
dispersion of the distant TFR was due entirely to galaxies with
non-relaxed kinematics, probably associated with mergers, as later
confirmed by \cite{puech08} (hereafter P08) as well as \cite{kassin07}
and \cite{covington09}. Restricting the smTFR to rotating disks (RDs),
for which the TFR can be confidently established, for the first time
P08 detected an evolution in zeropoint of the smTFR by
0.36$^{+0.21}_{-0.06}$ dex between z$\sim$0.6 and z=0. However,
z$\sim$0.6 galaxies have a larger gas fraction than their local
counterparts (i.e., $<$f$_{gas}>\sim$30\% at z$\sim$0.6-0.8), as
derived from the evolution of the gas-metallicity relation
\citep{rodrigues08}. This led P08 to suggest that the bTFR, as opposed
to the smTFR, might not be evolving.

In this paper, we take advantage of the representative IMAGES sample
of intermediate-mass galaxies at z$\sim$0.6 to establish the first
bTFR at high redshift, using the SK inversion method. The paper is
organized as follows: In Sect. 2, we present the sample and the data
used in this paper; In Sect. 3, we extend the previous analysis of the
smTFR by P08, obtaining new insight about its shift in zeropoint and
its scatter; In Sect. 4, we estimate the gas content of
intermediate-redshift galaxies and compare it to the stellar light
from young stars; In Sect. 5, we establish the first bTFR at
z$\sim$0.6, which is discussed in Sect. 6. Throughout, we adopt
$H_0=70$ km/s/Mpc, $\Omega _M=0.3$, and $\Omega _\Lambda=0.7$, and the
$AB$ magnitude system.

\section{Data \& sample}

\subsection{The sample}
We started from the 3D sample of P08. Galaxies were selected using
J-band absolute magnitudes as a proxy for stellar mass, such that $M_J
\leq -20.3$, which roughly corresponds to $M_{stellar}\geq 1.5\times
10^{10}M_\odot$, using a ``diet'' Salpeter IMF and \cite{bell03}
simplified recipes for deriving stellar mass from J-band luminosity.
Additional practical constraints were imposed on their rest-frame
[OII] equivalent width (i.e., $EW_0 \geq 15 \AA$) and redshift (i.e.,
$z$ between 0.4 and 0.75), by observing with the multi-integral-field
unit spectrograph FLAMES/GIRAFFE at the VLT. This provided a sample of
63 galaxies that represents the J-band luminosity function of galaxies
at these redshifts \citep{yang07}. In the following, we restrict our
study of the bTFR to galaxies within the CDFS (see Sect. 2.4). This
sample of 35 galaxies is still representative of the luminosity
function at z$\sim$0.6 \citep{yang07}.

NIR photometry was derived from public images of the respective fields
where galaxies were selected (i.e., CDFS, CFRS, and HDFS). Stellar
masses $M_{stellar}$ were estimated from $M_{stellar}/L_K$ ratios
using the method of \cite{bell03}, assuming a ``diet'' Salpeter IMF.
We estimated in P08 that this method, when applied to z$\sim$0.6
galaxies, provides us with estimates of the stellar mass with an
associated random uncertainty of 0.3 dex, and a systematic uncertainty
of -0.2 dex. We note that this comparison takes into account the
influence of possible secondary bursts of star formation
\citep{borch06}. In addition, the influence of TP-AGB stars on the
derivation of stellar masses could overestimate the stellar mass by up
to $\sim$0.14 dex \citep{maraston06,pozzetti07} in a systematic way,
but constant with redshift. In \cite{hammer09b}, it is shown that the
\cite{bell03} method, when applied to z$\sim$0.6 galaxies, provides us
with an upper limit to their stellar mass relative to the measurements
determined by other stellar-population-synthesis models and IMF
combinations, which is reflected by all systematic effects tending to
lower these estimates. Using these stellar masses to study the TFR is
therefore a quite conservative choice, leading to evolutionary trends
measured being perhaps lower than in reality. We refer the reader to
P08 and its Appendix for a clearer description of our adopted method,
as well as \cite{hammer09b}.

We note that the IMF used to derive stellar masses (i.e., the diet
Salpeter IMF following \citealt{bell03}) differs from the IMF used in
the next section to derive star formation rates (i.e., the
\emph{regular} Salpeter IMF). However, in \cite{hammer09b}, we argue
that stellar masses derived using \cite{bell03} recipes and a diet
Salpeter are roughly equivalent (in a one-to-one correlation and not
only in statistical sense) to stellar masses derived using the models
of \cite{bruzual03} and a \emph{regular} Salpeter IMF. Therefore, in
the following, we make no attempt to correct for this difference in
IMF, since the current generation of stellar population synthesis
models is simply not accurate enough for difference caused by these
two IMFs to be significant.

Compared to P08, we modified the classification of two galaxies in the
sample. J033245.11-274724.0 was studied in detail by \cite{hammer09},
who found in HST/ACS images from the UDF a dust-enshrouded disk that
had not previously been seen in GOODS images because of their
shallower depth (see a comparison between GOODS and UDF images in
their Fig. 1). Accounting for this disk, the dynamical axis was found
to be strongly misaligned with the morphological axis. \cite{hammer09}
argued that the morpho-kinematic properties of J033245.11-274724.0 can
only be reproduced by a major merger event. Similarly, \cite{puech09}
carried out a detailed study of J033241.88-274853.9, whose
morpho-kinematic properties also appeared to be far more closely
reproduced by a major merger model than that of a simple rotating
disk. Therefore, we shifted the classification of both objects from RD
to CK.

\subsection{Star formation rates}
\begin{figure}
\centering
\includegraphics[width=8.5cm]{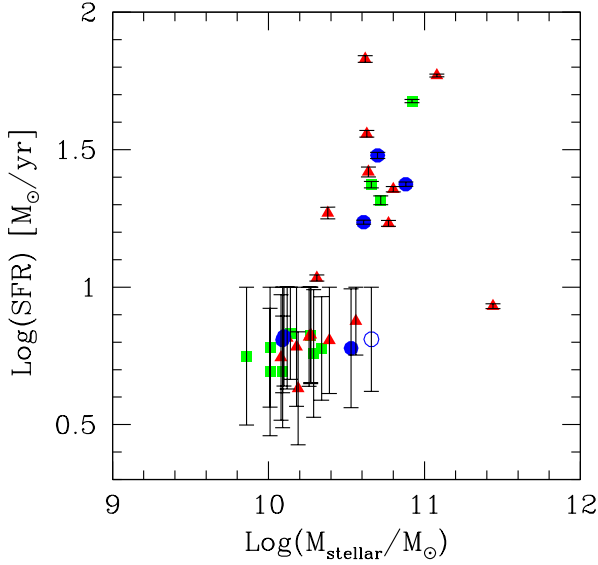}
\caption{SFR as a function of stellar mass in a subsample of 35 CDFS
  galaxies. RDs are shown as blue dots, PRs as green squares, and CKs
  as red triangles. The open blue circle corresponds to the RD+ galaxy
  for which the velocity measurement is more uncertain (see P08).
  Uncertainties in stellar mass are discussed in Sect. 2.1, and are
  not shown for reasons of clarity.}
\label{sfr}
\end{figure}

We estimated the total SFR$_{tot}$ of each galaxy, following
\cite{puech07b}. Briefly, SFR$_{tot}$ was taken to be the sum of
SFR$_{UV}$, derived from the 2800\AA~luminosity $L_{2800}$, and
SFR$_{IR}$, which was derived from Spitzer/MIPS photometry at 24$\mu
m$ using the \cite{chary01} calibration between rest-frame 15$\mu m$
flux and total IR luminosity $L_{IR}$. To convert both $L_{IR}$ and
$L_{2800}$ into SFRs, we used the calibrations of \cite{kennicutt98},
which rely on a Salpeter IMF. Uncertainties were estimated by
propagating the flux uncertainty measurement using the SFR
calibrations. About half of the galaxies in the sample were not
detected by MIPS. In this case, we derived an upper limit to
SFR$_{IR}$ as a function of redshift using Fig. 9 of \cite{lefloch05}.
We assumed that SFR$_{tot}$ is the mean of this limit and SFR$_{UV}$.
The corresponding uncertainty in this case is half the difference
between SFR$_{UV}$ and the limit to SFR$_{IR}$, which infers a mean
relative uncertainty of 50\% for these objects (see Fig. \ref{sfr} and
Table \ref{tab}).

\subsection{Gas mass estimates}
To estimate the mass of gas within each galaxy, we used the SK
relation between SFR and gas densities of \cite{kennicutt89}, which is
given by $\Sigma _{SFR}=2.5\times 10^{-4} \Sigma _{gas}^{1.4}$
M$_\odot$ /yr/kpc$^{2}$. \cite{bouche07} derived a new calibration of
the SK law, based on distant sources (z$\sim$2-3). They found that the
SK law holds up to z$\sim$2.5, but has a quite different power-law
index (i.e., 1.7 instead of 1.4). For the CDFS sample, we find that
$\Sigma _{gas}$ ranges from 0.004 to 0.46 M$_\odot$/yr/kpc$^2$. In
this regime, the two SK laws are relatively similar, the discrepancy
between the two relations being caused mostly by the higher
star-formation densities found in z$\sim$2 starbursts (see Fig. 3 of
\citealt{bouche07}). This range corresponds to the star-formation
densities of local (U)LIRGs, starbursts, and star-forming disks
\citep{bouche07}. Finally, at z=0.24, \cite{lah07} performed direct HI
measurements of distant galaxies, and found that there is no evolution
in the relation between HI mass and SFR. Therefore, we conclude that
it is more appropriate to rely on the local SK law rather than the one
at high-z to derive gas masses in a sample of z$\sim$0.6 galaxies.

\begin{figure}
\centering
\includegraphics[width=8.5cm]{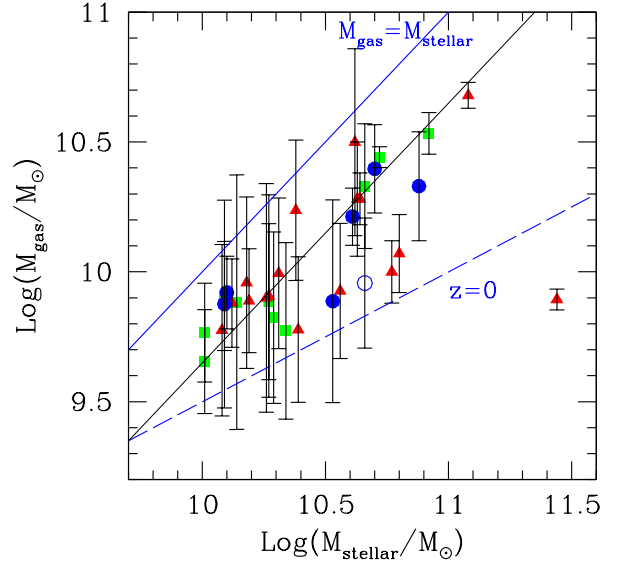}
\caption{Gas mass as a function of stellar mass in a subsample of
  35 CDFS galaxies. RDs are shown as blue dots, PRs as green squares,
  and CKs as red triangles. The open blue circle corresponds to the
  RD+ galaxy for which the velocity measurement is more uncertain (see
  P08). Uncertainties in stellar mass are discussed in Sect. 2.1, and
  are not shown for ease of clarity. The blue line represents equality
  between the stellar and gas masses, while the blue dashed line
  represents the local relation between these two quantities given by
  \cite{schim08}. The median gas fraction is found to be 31\%, or
  M$_{gas}$/M$_{stellar}$=0.45, which is represented as a thin black
  line.}
\label{mgas}
\end{figure}

To do this, we first derived the SFR density $\Sigma
_{SFR}=SFR_{total}/\pi R_{gas}^2$, which was then converted into a gas
density $\Sigma _{gas}$ using the SK law. Gas masses were derived as
$M_{gas}=\Sigma _{gas} \times \pi R_{gas}^2$. The same radius
$R_{gas}$, which is defined in the next section, is assumed when
normalizing both densities, following \cite{kennicutt89}.
Uncertainties in $M_{gas}$ were derived using standard methods of
error propagation (see Table \ref{tab} and Fig. \ref{mgas}). It is
noteworthy that gas masses derived this way are independent of the
IMF, since the SK law and SFRs are derived using the same Salpeter
IMF\footnote{The IMF affects the SFR calibration in such a way that
  both the SFR and the numerical factors in the SK relation change and
  cancel eachother. It is possible to derive a direct relation between
  UV and IR luminosities and gas densities, in a similar way to
  \cite{erb06} using the H$\alpha$ luminosity, therefore eliminating
  the SFR density, which is the IMF-dependant quantity.}. Gas masses
and gas fractions in z$\sim$0.6 galaxies are discussed further by
\cite{hammer09b}. Of particular interest here, is that the gas masses
derived by inverting the SK law lead to a median gas fraction of 31\%
in z$\sim$0.6 galaxies. We note that gas fractions, in contrast to gas
masses, are not IMF independent. Strikingly, the same value is found
using a completely different data set and methodology (i.e., based on
the evolution of the mass-metallicity relation derived from FORS2
long-slit spectroscopy, see \citealt{rodrigues08}). It is therefore
very unlikely that the gas fractions derived at z$\sim$0.6 could be
affected by any systematic effect, which makes us quite confident in
our estimates of gas masses.

\subsection{[OII] gas radius estimates}
To convert from density to mass, one needs to define and measure a gas
radius $R_{gas}$. To do this, we first constructed rest-frame UV
images by summing observed B and V bands. Given that not all galaxies
in the sample have homogeneous imaging (see \citealt{neichel08}), we
decided to restrict the study of the bTFR to galaxies with the highest
quality images. Therefore, we restricted the study of the bTFR to the
subsample of 35 galaxies lying in the CDFS with HST/ACS images, which
is still representative of the luminosity function at z$\sim$0.6
\citep{yang07}. This allowed us to limit uncertainties in the
derivation of gaseous radii and therefore, gas masses.

From rest-frame UV images, we derived for each galaxy an axis ratio
$b/a$ and a $PA$ using SExtractor \citep{bertin96}. For each galaxy,
these two parameters were used to generate a set of images with flat
ellipses of increasing radii using the IDL procedure DIST\_ELLIPSE. To
account for the relatively coarse spatial sampling of the GIRAFFE IFU
(0.52 arcsec/pix, which roughly corresponds to 3.5 kpc at z$\sim$0.6),
these high-resolution ellipses were then rebinned to the GIRAFFE pixel
scale after determining as accurately as possible the position of the
GIRAFFE IFU on the UV image (see, e.g., \citealt{puech07b}). These
simulated [OII] GIRAFFE images were renormalized in terms of flux
using the observed IFU-integrated [OII] value, and pixels with a
resulting flux lower than the minimal [OII] flux detected within the
GIRAFFE IFU for a given galaxy were disregarded, to account for
GIRAFFE IFU pixels being selected in terms of signal-to-noise ratio
(i.e., S/N$\geq$3, see \citealt{flores06}). By identifying the
simulated [OII] map that most closely reproduces the observed one, it
is in principle possible to retrieve the underlying [OII] total-light
radius. However, the exact position of the IFU, as well as the precise
value of the seeing during the observations are subject to
uncertainties (e.g., \citealt{puech07b,puech08}) that prevent us from
being able to determine the exact position of the IFU grid. The
relatively coarse spatial resolution of the GIRAFFE IFU also makes it
relatively difficult to retrieve precisely the underlying [OII] flux
distribution. To mitigate the uncertainty associated with these
effects, we did not use a classical chi-square minimization between
the simulated and observed [OII] maps, but used a spatial
cross-correlation to determine, for each galaxy, the subset of
simulated [OII] maps that maximizes the number of pixels illuminated
in both the observed and simulated [OII] maps. This subset of
simulated maps corresponds to a range of ellipse radii that are
deconvolved from the GIRAFFE coarse spatial sampling. We assumed that
the mean value of this range is a measure of the total [OII] radius
$R_{[OII]}$, whose error-bars are equal to half this range. Finally,
$R_{[OII]}$ were quadratically deconvolved from a 0.8 arcsec mean
seeing disk. The maximal value of the ratio of the errorbars to the
corresponding radii is found to be 51\%, while their distribution
shows an average of 17\% and a 1-$\sigma$ spread of 12\%. The derived
radii and errorbars in kpc are shown in Fig. \ref{figrad} and listed
in Table \ref{tab}.

\begin{sidewaystable*}[p]
\centering
\caption{Principle properties of the sample of 35 galaxies used in
  this study, in order of increasing $RA$ (see Sect. 2.5). From left
  to right: IAU ID, redshift (from \citealt{ravi07}), dynamical class
  (from \citealt{yang07}), rotation velocity and associated
  uncertainty (from P08), stellar mass $\log{(M_{stellar}/M_\odot)}$
  (from P08), SFR$_{UV}$ (in M$_\odot$/yr), SFR$_{IR}$ (in
  M$_\odot$/yr), total SFR and associated uncertainty $\Delta$SFR (in
  M$_\odot$/yr), rest-frame UV half-light radius R$_{UV}$ and
  associated uncertainty $\Delta$R$_{UV}$ (in kpc), [OII] total radius
  R$_{[OII]}$ and associated uncertainty $\Delta$R$_{[OII]}$ (in kpc),
  and mass of gas $\log{(M_{gas}/M_\odot)}$ with associated
  uncertainty $\Delta \log{(M_{stellar}/M_\odot)}$. }
\begin{tabular}{cccccccccccccccc}\hline
IAU ID & $z$ & D.C. & $V_{flat}$ & $\Delta V_{flat}$ &
$\log{(M_{stellar}/M_\odot)}$ & SFR$_{UV}$ & SFR$_{IR}$ & SFR &
$\Delta$SFR & R$_{UV}$& $\Delta$R$_{UV}$ & R$_{[OII]}$ &
$\Delta$R$_{[OII]}$ & $\log{(M_{gas}/M_\odot)}$ & $\Delta
\log{(M_{gas}/M_\odot)}$\\\hline

J033210.25-274819.5 & 0.6100 &  PR  &  150 &  26 & 10.29 & 1.68 & --- & 5.7  & 4.1 & 4.46 & 0.10 & 9.80  & 1.83 & 9.82 & 0.33\\
J033210.76-274234.6 & 0.4180 &  CK  &  550 & 123 & 11.44 & 3.25 & 5.27 & 8.5  & 0.2 & 4.18 & 0.08 & 5.65  & 0.18 & 9.89 & 0.04\\
J033212.39-274353.6 & 0.4230 &  RD  &  180 &  22 & 10.61 & 1.04 & 16.18& 17.2 & 0.3 & 6.27 & 0.08 & 8.25  & 0.94 & 10.21& 0.11\\
J033213.06-274204.8 & 0.4220 &  CK  &  130 &  22 & 10.19 & 1.69 & --- & 4.3  & 2.6 & 7.29 & 0.09 & 18.35 & 0.86 & 9.89 & 0.20\\
J033214.97-275005.5 & 0.6680 &  PR  &  190 &  90 & 10.92 & 7.01 & 40.46& 47.5 & 0.7 & 6.42 & 0.12 & 9.19  & 0.69 & 10.53& 0.08\\
J033217.62-274257.4 & 0.6470 &  CK  &  250 &  43 & 10.38 & 1.89 & 16.70& 18.6 & 0.9 & 3.16 & 0.11 & 11.94 & 2.86 & 10.24& 0.27\\
J033219.32-274514.0 & 0.7250 &  CK  &  270 &  29 & 10.39 & 2.81 & --- & 6.4  & 3.6 & 3.74 & 0.12 & 6.96  & 1.02 & 9.78 & 0.28\\
J033219.61-274831.0 & 0.6710 &  PR  &  190 &  33 & 10.27 & 3.36 & --- & 6.7  & 3.3 & 2.93 & 0.11 & 10.41 & 1.99 & 9.89 & 0.30\\
J033219.68-275023.6 & 0.5610 &  RD  &  230 &  33 & 10.88 & 5.06 & 18.61& 23.7 & 0.4 & 6.75 & 0.10 & 12.71 & 2.31 & 10.33& 0.21\\
J033220.48-275143.9 & 0.6790 &  CK  &   70 &  24 & 10.18 & 2.15 & --- & 6.1  & 3.9 & 2.54 & 0.11 & 15.73 & 3.19 & 9.96 & 0.33\\
J033224.60-274428.1 & 0.5380 &  CK  &   90 &  27 & 10.08 & 1.72 & --- & 5.6  & 3.8 & 3.88 & 0.10 & 8.42  & 1.55 & 9.78 & 0.33\\
J033225.26-274524.0 & 0.6660 &  CK  &   80 &  26 & 10.56 & 5.05 & --- & 7.5  & 2.5 & 2.92 & 0.11 & 10.57 & 1.74 & 9.93 & 0.26\\
J033226.23-274222.8 & 0.6679 &  PR  &  200 &  24 & 10.72 & 3.17 & 17.53& 20.7 & 0.8 & 12.52& 0.13 & 13.84 & 0.53 & 10.44& 0.04\\
J033227.07-274404.7 & 0.7390 &  CK  &  110 &  21 & 10.26 & 3.20 & --- & 6.6  & 3.4 & 4.11 & 0.11 & 11.18 & 4.61 & 9.90 & 0.44\\
J033228.48-274826.6 & 0.6697 &  CK  &  130 &  66 & 10.63 & 1.47 & 34.61& 36.1 & 1.0 & 1.62 & 0.11 & 6.21  & 1.15 & 10.28& 0.22\\
J033230.43-275304.0 & 0.6460 &  CK  &  380 &  29 & 10.64 & 1.56 & 24.67& 26.2 & 1.1 & 4.88 & 0.14 & 9.06  & 0.59 & 10.28& 0.10\\
J033230.57-274518.2 & 0.6810 &  CK  &   80 &  46 & 11.08 & 11.24& 47.57& 58.8 & 0.7 & 8.85 & 0.11 & 10.28 & 0.52 & 10.68& 0.05\\
J033230.78-275455.0 & 0.6870 &  RD+ &  200 &  43 & 10.66 & 2.92 & --- & 6.5  & 3.5 & 7.61 & 0.11 & 7.93  & 1.71 & 9.96 & 0.25\\
J033231.58-274121.6 & 0.7047 &  RD  &  140 &  41 & 10.10 & 3.22 & --- & 6.6  & 3.4 & 6.28 & 0.12 & 12.14 & 0.08 & 9.92 & 0.14\\
J033232.96-274106.8 & 0.4690 &  PR  &  210 & 117 & 10.01 & 1.45 & --- & 4.9  & 3.5 & 1.24 & 0.09 & 6.03  & 0.15 & 9.66 & 0.20\\
J033233.90-274237.9 & 0.6190 &  PR  &  200 & 107 & 10.66 & 5.69 & 17.88& 23.6 & 0.6 & 2.94 & 0.10 & 12.92 & 2.75 & 10.33& 0.24\\
J033234.04-275009.7 & 0.7030 &  RD  &  160 &  29 & 10.09 & 2.86 & --- & 6.4  & 3.6 & 4.26 & 0.11 & 10.52 & 3.45 & 9.88 & 0.40\\
J033234.12-273953.5 & 0.6280 &  CK  &  110 &  44 & 99.99 & 3.26 & 25.82& 29.1 & 1.4 & 3.73 & 0.12 & 6.37  & 1.33 & 10.25& 0.22\\
J033237.54-274838.9 & 0.6650 &  RD  &  230 &  71 & 10.70 & 7.88 & 22.26& 30.2 & 0.8 & 6.53 & 0.11 & 11.50 & 1.65 & 10.40& 0.17\\
J033238.60-274631.4 & 0.6220 &  RD  &  210 &  33 & 10.53 & 2.13 & --- & 6.0  & 3.9 & 5.91 & 0.11 & 11.99 & 3.47 & 9.89 & 0.39\\
J033239.04-274132.4 & 0.7330 &  PR  &  130 &  39 & 10.14 & 3.60 & --- & 6.8  & 3.2 & 2.84 & 0.11 & 10.09 & 5.14 & 9.88 & 0.49\\
J033239.72-275154.7 & 0.4160 &  CK  &   30 &  36 & 10.31 & 4.19 & 6.61 & 10.8 & 0.3 & 3.06 & 0.08 & 8.87  & 2.43 & 10.00& 0.29\\
J033240.04-274418.6 & 0.5223 &  CK  &  470 & 214 & 10.77 & 1.67 & 15.38& 17.1 & 0.4 & 1.95 & 0.09 & 5.11  & 0.43 & 10.00& 0.12\\
J033241.88-274853.9 & 0.6680 &  RD+ &  120 &  27 & 10.27 & 3.41 & --- & 6.7  & 3.3 & 3.09 & 0.11 & 11.29 & 3.59 & 9.91 & 0.39\\
J033243.62-275232.6 & 0.6800 &  PR  &   60 &  23 &  9.86 & 1.23 & --- & 5.6  & 4.4 & 4.56 & 0.15 & 17.04 & 0.05 & 9.95 & 0.20\\
J033244.20-274733.5 & 0.7365 &  CK  &  170 &  40 & 10.62 & 5.77 & 61.77& 67.5 & 1.9 & 1.49 & 0.11 & 6.84  & 2.54 & 10.50& 0.36\\
J033245.11-274724.0 & 0.4360 &  RD  &  270 &  44 & 10.80 & 1.92 & 20.77& 22.7 & 0.5 & 1.63 & 0.09 & 4.75  & 0.56 & 10.07& 0.15\\
J033248.28-275028.9 & 0.4462 &  PR  &  110 &  22 & 10.09 & 1.98 & --- & 4.9  & 2.9 & 8.77 & 0.12 & 14.41 & 0.94 & 9.91 & 0.21\\
J033249.53-274630.0 & 0.5230 &  PR  &  150 &  39 & 10.34 & 2.74 & --- & 6.0  & 3.3 & 2.98 & 0.10 & 7.58  & 1.72 & 9.77 & 0.34\\
J033250.24-274538.9 & 0.7318 &  CK  &  240 &  67 & 10.12 & 3.06 & --- & 6.5  & 3.5 & 5.51 & 0.13 & 6.88  & 0.26 & 9.88 & 0.17\\
J033250.53-274800.7 & 0.7370 &  PR  &  110 &  26 & 10.01 & 2.10 & --- & 6.1  & 4.0 & 3.83 & 0.12 & 6.85  & 0.18 & 9.77 & 0.19\\\hline
\end{tabular}
\label{tab}
\end{sidewaystable*}

\section{Evidence that major mergers are responsible for the large scatter in the distant TFR}

\subsection{The z$\sim$0.6 stellar-mass TFR}
\begin{figure*}
\centering
\includegraphics[width=8.5cm]{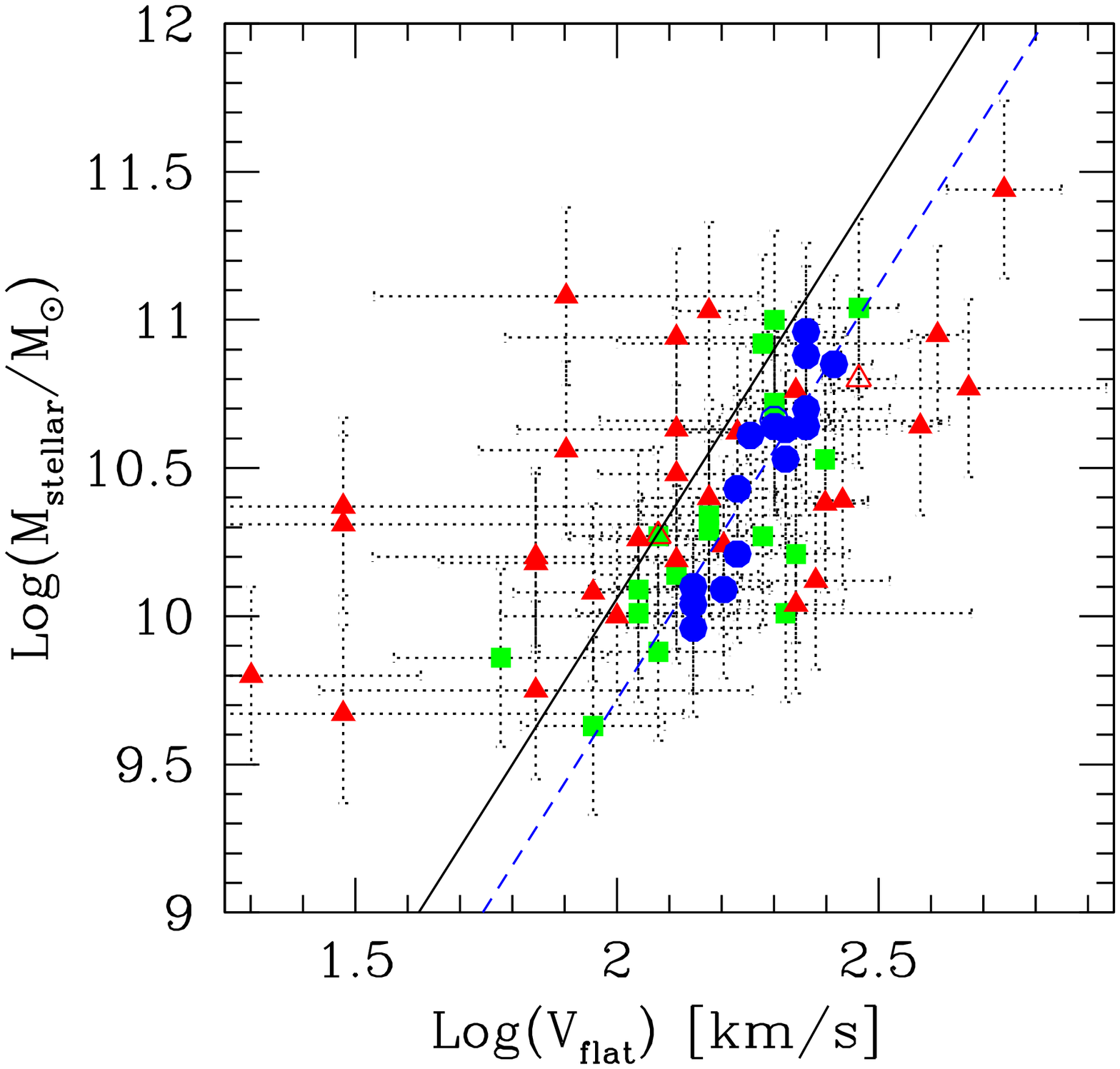}
\includegraphics[width=8.5cm]{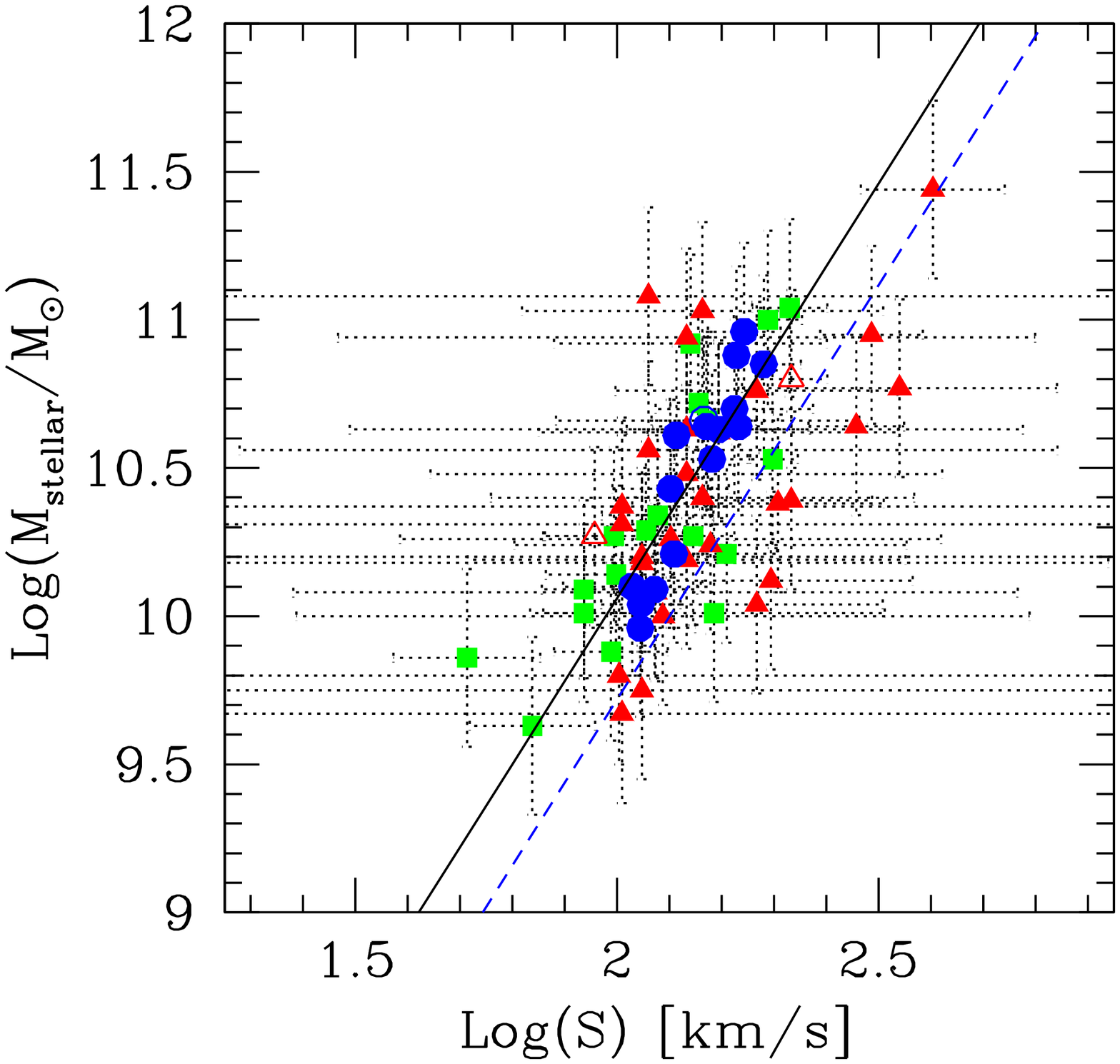}
\caption{\emph{left:} Evolution of the stellar-mass TFR in the sample
  of 64 galaxies of P08 (RDs are shown as blue dots, PRs as green
  squares, and CKs as red triangles). The open blue circle correspond
  to the RD+ galaxies for which the velocity measurement is more
  uncertain, see P08). The two open red triangles correspond to the
  two galaxies that were changed from RD to CK compared to P08 (see
  Sect. 2.1). The black line is the local smTFR, while the blue
  dash-line represent a linear fit to the $z\sim0.6$ smTFR.
  \emph{Left:} Same relation but using $S$. The black and blue-dashed
  lines are the same as in the left panel.}
\label{smtf}
\end{figure*}

The smTFR in the IMAGES sample was initially derived in P08 (see their
Appendix). We show in Fig. \ref{smtf} (see left panel) a revised
version of this relation, in which the classification of two objects
has changed (see Sect. 2). We also show here all galaxies, including
non-relaxed ones. The dispersion in this relation was shown to be
caused entirely by non-relaxed systems by P08 (see next section).
Maintaining the slope at its local value, we found a shift in the
smTFR zeropoint of 0.34$^{+0.21}_{-0.06}$ dex in stellar mass, or,
equivalently, of $\sim$0.12 dex in velocity, between $z\sim0.6$ and
$z=0$, consistent with the shift derived by P08. The errorbars account
for possible systematic effects, which were quantified in detail by
P08. Most of them are associated with the estimation of stellar mass.
As we noted in Sect. 2.1, all these effects would tend to decrease the
estimates of the stellar mass in z$\sim$0.6 galaxies, since we
``maximized'' them by adopting the \cite{bell03} method. The only
systematic effects that could reduce the shift in zeropoint are
relatively insignificant (i.e., -0.06 dex), and are associated with
the derivation of rotation velocities (see P08). In P08, we found that
the shift of the smTFR with redshift cannot be interpreted as a pure
evolution along the velocity axis, and that only half, at most, could
be accounted for by a velocity shift caused by gas accretion within
the optical radius. We then concluded that most of the shift in the
smTFR reflects an increase in stellar mass in rotating disks over the
past 6 Gyr.

\cite{koen09} performed a permutation test to demonstrate that the
distant K-band TFR found by P08 was offset significantly from the
local relation. We repeated their analysis for the smTFR and found a
probability $\ll$1\% that the two relations have identical slope and
zeropoint. To test the reliability of this result in terms of
systematic uncertainties, we shifted the distant smTFR in steps of
0.01 dex in stellar mass toward the local relation and repeated the
permutation test. We found that for this probability to exceed 10\% we
would need to bias the stellar mass in the distant sample by +0.28
dex, which exceeds systematic uncertainties (see above). This means
that, within the systematic effects on the zeropoint shift identified
by P08, the two relations differ significantly, which secures the
results of P08.

\subsection{Origin of the scatter in the z$\sim$0.6 smTFR}
In \cite{flores06}, we showed that the dispersion in the z$\sim$0.6
smTFR was caused by non-relaxed galaxies. This result was confirmed by
P08, who showed that the dispersion in the distant smTFR, once
restricted to RDs (i.e., 0.12 dex), was similar to that in the local
relation, with 0.15 dex. The galaxies contributing to the scatter in
the distant smTFR exhibit strong perturbations in their kinematic maps
(e.g., \citealt{yang07}), which led us to claim that most of them were
probably associated with major mergers (see P08). This interpretation
is also supported by the large scatter seen in the specific angular
momentum versus rotation velocity plane, which could be produced by
these events, as demonstrated by \cite{puech07}.

Interestingly, \cite{weiner06} defined a new kinematic tracer
$S=\sqrt{0.5\times V_{rot}^2+\sigma ^2}$ that combines rotation
velocity and velocity dispersion. They found that velocity dispersion
is an important component of the dynamical support in
intermediate-redshift galaxies, especially for morphologically compact
or disturbed systems. Their study confirmed the earlier finding of
\cite{puech06} that the majority of compact galaxies have motions in
which dispersion plays an important role (see also \citealt{puech07}).
\cite{kassin07} later showed that using $S$ in the smTFR leads to a
significant reduction in its scatter. In addition, \cite{covington09}
showed, using hydro-dynamical simulations of major mergers, that this
reduction in scatter is consistent with a (partial) transfer of energy
between the kinetic energy associated with bulk motions to that
associated with random motions. This transfer is driven by shocks and
collisions generated during the merger events (e.g.,
\citealt{montero06}).

That the ``merger-hypothesis'' is the underlying cause of the large
scatter measured in the distant TFR can be tested by searching for
specific objects and verifying whether this transfer of energy between
bulk and random motions driven by interaction-driven shocks really
occurs. \cite{peirani09} studied one CK galaxy of the sample
(J033239.72-275154.7) in detail and they demonstrated that both its
morphology and kinematics were reproduced well by a 1:3 major merger
where the companion galaxy had been in a retrograde parabolic orbit
and there had been an inclination angle of 15 degrees between the two
orbital planes. This object is a perfect test-bed for the present
discussion, since it is measured to have $\log{V_{flat}}$=1.48 km/s
and $M_{stellar}=10.31$ M$_\odot$, i.e., to be among the CK galaxies
in the distant smTFR that are the most deficient in rotation velocity
(see left panel of Fig. \ref{smtf}). \cite{peirani09} showed that,
because of the retrograde nature of interaction, the host galaxy can
lose angular momentum to its companion, resulting in the deceleration
of the main progenitor. This explains why the resulting merger remnant
appears to have systematically smaller velocities in the smTFR than
other objects. Shocks in the gaseous phase generated during the
interaction produce an increase in the velocity dispersion, similar to
that suggested by \cite{covington09} by their merger simulations (see
their Fig. 1). If the scatter in the smTFR is really caused by a
transfer of energy between bulk and random motions driven by merger
events, then this object should fall back onto the TFR, once its
velocity dispersion is taken into account.

To test this, we derived $S$ in the IMAGES sample using
$V_{rot}=V_{flat}$ and $\sigma = \sigma _{disk}$, as defined by
\cite{puech07}. We show the resulting $S$-TFR in Fig. \ref{smtf} (see
right panel). Using $S$ results in the shifting of all RDs back onto
the local smTFR, so we propose to use the local slope in all future
investigations. We also note that several CK galaxies lie at
relatively high $S$ compared to the bulk of the sample. These galaxies
were found to have large rotation velocity in the conventional smTFR.
However, these galaxies do not seem to be present in the
\cite{kassin07} smTFR, although \cite{covington09} managed to produce
some of these galaxies in their merger simulations (see their Fig. 3).
These differences seem to be related to different selection criteria
involving inclination. We did not apply any selection of this type
(see P08), while \cite{kassin07} removed galaxies with low
inclinations (i.e., $inc \leq 30$deg) because of the larger associated
uncertainty, and those with high inclinations (i.e., $inc \geq
70$deg), because of the effect of dust (see also
\citealt{covington09}).

The total scatter in the resulting $S$-TFR is significantly lower than
in the smTFR, since it decreases from 0.63 to 0.34 dex. This residual
scatter is still a factor of two larger than the residual dispersion
in the local relation, which is $\sigma _{res}$=0.15 dex (in mass).
Looking at the dispersion in more detail, the scatter remains roughly
constant in RDs and PRs (from 0.12 to 0.14 dex and 0.32 to 0.29 dex,
respectively), but strongly decreases from 0.83 to 0.41 dex among CKs.
That kinematically disturbed galaxies show a smaller but still large
scatter compared to the local relation suggests that part of the
remaining discrepancy is related to the greater uncertainty in
rotation velocity of CKs (see P08). The median uncertainty in the
rotation velocity of the sample is found to be 0.12 dex, which
translates into $\sim$0.3 dex in stellar mass, i.e., the same order of
magnitude as the residual scatter in the distant $S$-TFR.

As one can see in the right panel of Fig. \ref{smtf}, accounting for
velocity dispersion brings the merger-test J033239.72-275154.7 back
onto the $S$-TFR. This is the first direct observational evidence of
the transfer of energy described by \cite{covington09}. As shown
above, this process can account for most of the scatter in the smTFR.
Given the still relatively large uncertainties, we cannot exclude
other process(es) playing a role in producing the scatter in the
distant relation, but these results clearly indicate that this
transfer of energy is the main underlying cause, and that mergers are
the main cause of the large scatter seen in the z$\sim$0.6 TFR, as
initially argued by \cite{flores06} (see also \citealt{kannappan04}).

\section{Gas content of intermediate-mass galaxies}

\subsection{Comparison between gas and stellar extents}
\begin{figure}
\centering \includegraphics[width=8.7cm]{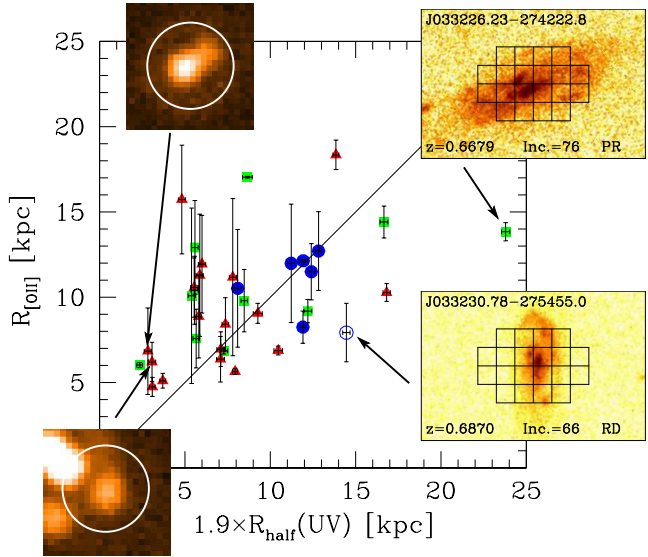}
\caption{Comparison between the [OII] and UV radii for the CDFS
  sub-sample of galaxies. The two insets on the left (30-arcsec wide)
  show MIPS imaging at 24 $\mu m$ of J033228.48-274826.6 (upper one)
  and J033244.20-274733.5, while the two insets on the right show
  z-band HST/ACS images of J033230.78-275455.0 and J033226.23-274222.8
  superimposed with the GIRAFFE IFU grid. RDs are shown as blue dots,
  PRs as green squares, and CKs as red triangles. The open blue circle
  corresponds to the RD+ galaxy for which the velocity measurement is
  more uncertain (see P08).}
\label{figrad}
\end{figure}

We show the total-light [OII] radius $R_{[OII]}$ as a function of 1.9
times the UV half-light radius $R_{UV}$ in Fig. \ref{figrad}. For a
thin exponential disk, 1.9 times the half-light radius is equal to the
optical radius of the galaxy \citep{persic91}, which provides us with
a useful comparison point\footnote{For a thin exponential disk, this
  radius equals 3.2 times the disk scale-length, which defines the total
  stellar-light radius, and is statistically equivalent to the
  isophotal radius R$_{25}$, see \cite{persic91}.}. Galaxies that have
a $R_{[OII]}$ significantly lower than 1.9$\times R_{UV}$ are galaxies
for which the UV light extents farther than the IFU FoV (3$\times$2
arcsec$^2$). Two examples of these galaxies can be seen on the right
side of Fig. \ref{figrad}. For these galaxies, it is clear that
assuming $R_{[OII]}$ to be a measure of the gaseous extent would lead
to underestimating the gas mass. Therefore, we defined
$R_{gas}=MAX(R_{[OII]},1.9\times R_{UV})$ to be a conservative measure
of the gaseous extent. We note that for galaxies with non-relaxed
kinematics, this might still underestimate the gas radius. The
uncertainties correspond to the respective uncertainties in
$R_{[OII]}$ or 1.9$\times R_{UV}$ depending on which radius was the
largest for a given galaxy.

Most RDs (see blue dots) fall relatively close to the equality line,
which instills confidence in the estimate of $R_{gas}$ in dynamically
relaxed systems. To first order, one can assume that, at least on
large spatial scales and for relaxed systems, the spatial distribution
of UV light emitted by young OB stars is correlated with the emission
from the [OII] ionized gas detected by GIRAFFE. However, in more
disturbed galaxies, there is a clear trend in which the gas is more
extended than the UV light. We find a median [OII]-to-UV radius ratio
of 1.3$\pm 0.16$ (1-$\sigma$ bootstrapped uncertainty), which is
consistent with the mean ratio of $\sim$1.2 of emission line to B-band
scale-lengths found by \cite{bamford07} in a sample of field galaxies
at similar redshifts. Using a Student t-test, the probability that
this median ratio is equal to one (i.e., that the total [OII] and UV
radii are statistically equivalent) is found to be 6\%. This trend is
even more pronounced for more compact systems: galaxies with a UV
radius lower or equal to the median in the sample have a median ratio
of $\sim$1.7, while those having a UV radius larger than the median
value, have a ratio $\sim$0.9. We also examined the GOODS/MIPS (DR3)
images of the four galaxies with a UV radius smaller that one GIRAFFE
pixel. All were detected, although only three at a significantly high
level for their flux to be measurable. Interestingly, two of them were
resolved by MIPS (see the two insets on the left side of Fig.
\ref{figrad}), which confirms that the gas content can be
significantly more extended than the UV stellar light.

\subsection{What causes the large gaseous extent in z$\sim$0.6 galaxies?}
It is interesting to investigate what causes this discrepancy between
the extents of the ionized gas and the UV ionizing stellar light from
OB stars. Most of the ionizing stellar light is assumed to be emitted
by stars earlier than B2 \citep{stromberg39}, so to first order, one
would expect conversely that the ionized gas is confined to within the
UV radius. What mechanisms can explain our actual findings?

Supernovae (SN) and AGN feedback are two mechanisms that are known to
be able to drive outflows in galaxies. If these mechanisms were
responsible for ejecting ionized gas to the UV radius, we should
detect offsets between absorption and emission lines in a large
fraction of galaxies in the sample. For 20 galaxies in the GIRAFFE
sample, we were able to retrieve FORS2 integrated spectra (see
\citealt{rodrigues08}) that allowed us to measure absorption lines.
Among these galaxies, we found systematic shifts in only three or
possibly four of them (at a $\sim$100km/s level in
J033224.60-274428.1, J033225.26-274524.0, J033214.97-275005.5, and
possibly in J033210.76-274234.6). Hence, we conclude that outflows,
regardless of their powering mechanism, cannot explain why gas extends
farther out than the UV stellar light.

If ionized gas is not pushed out of the UV disk by outflows, then the
gas must be ionized by a mechanism other than the radiation of OB
stars. Interestingly, for one object in the sample, \cite{puech09}
observed a region that is completely devoid of stars, where ionized
gas was detected. They concluded that this gas was ionized by shocks
induced by a major merger. These interaction-driven shocks could be an
important mean of ionizing gas in z$\sim$0.6 galaxies.
\cite{hammer09b} showed that the morpho-kinematics of most galaxies in
the sample can be reproduced by the simulations of major mergers. We
note that gas in all galaxies in the sample have lower V/$\sigma$
ratios than their local counterparts, which might be a dynamical
signature of these shocks \citep{puech07}. In particular, most PRs
show off-center velocity dispersion peaks (see \citealt{yang07}),
which strengthens this interpretation.

\section{The baryonic Tully-Fisher relation at z$\sim$0.6}

\subsection{The local relation}
We used the local sample gathered by \cite{mcgaugh05} to derive our
local reference bTFR. \cite{mcgaugh05} used $V_{flat}$ and (amongst
others) the \cite{bell03} method to estimate stellar mass, so their
sample is particularly well suited to compare with our high-z bTFR.
However, they used the B-band luminosity to derive stellar mass, which
is a less accurate tracer than K-band luminosities. Hence, we
rederived stellar mass from K-band luminosities by cross-correlating
the \cite{mcgaugh05} sample with the 2MASS database. Stellar masses
were derived following the method outlined in P08 for their local
sample, i.e., by accounting for k-corrections and extinction. We found
that $\log{(M_{baryonic}/M_\odot)}=2.10 \pm 0.42 + (3.74\pm0.20)\times
\log{(V_{flat})}$ has a residual scatter of $\sigma$=0.25 dex. In the
following, we checked that by instead using the \cite{mcgaugh05}
stellar masses derived from the B-band luminosity (see his Table 2
with \ensuremath{{\cal P}}=1) would not change significantly our
results.

\cite{stark09} established a new calibration of the local bTFR using
gas-dominated galaxies, which has the advantage of being less
sensitive to a given estimator of the stellar mass. However,
\cite{stark09} averaged the bTFR produced by various stellar mass
estimators. Therefore, we chose to keep our own derived relation,
which has the advantage of resulting from the same methodology used to
derive the smTFR, both at low and high redshift, as well as the bTFR
at high redshift, which minimizes possible systematic effects. We
nevertheless note that our slope and zeropoint are consistent with
their calibration. In contrast to the smTFR, whose calibration depends
on the galaxy populations analyzed, the bTFR is found to be
independent of galaxy type \citep{stark09}. We therefore do not expect
any bias to be caused by the choice of a particular local sample and
its level of representation in the local population.

\subsection{The baryonic TFR at z$\sim$0.6}
\begin{figure}
\centering
\includegraphics[width=9.7cm]{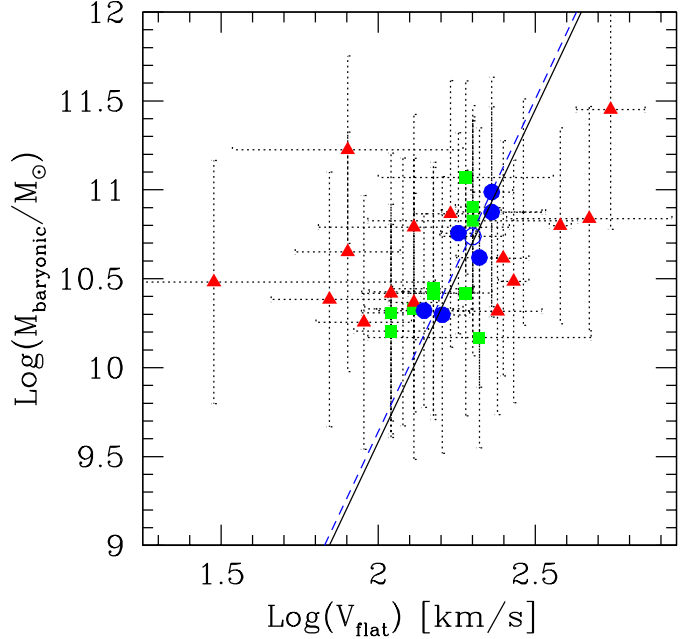}
\caption{Evolution of the baryonic TFR in the CDFS subsample (see
  text). The black line is the local bTFR of \cite{mcgaugh05}, while
  the blue-dashed line is the fit to the distant relation. All panels
  have similar limits to ease the comparison between the different
  relations}
\label{btf}
\end{figure}

We show the first bTFR obtained at high redshift so far, in Fig.
\ref{btf}. Overall, the bTFR has the same structure as the smTFR, with
non-relaxed galaxies associated with the greatest amount of
dispersion. Holding the slope constant, we found a distant zeropoint
of $2.16\pm 0.07$, i.e., a shift of +0.06 dex in baryonic mass between
the z=0 bTFR and z$\sim$0.6 RDs. This shift translates into a -0.02
dex shift in rotation velocity. The scatter in the distant bTFR for
RDs is found to be roughly similar to that in the local bTFR, with
0.14 dex, which suggests no evolution in slope. In comparison with the
distant smTFR, the scatter in RDs is found to be roughly equivalent,
with 0.14 dex in the bTFR instead of 0.12 dex in the smTFR (i.e.,
bootstrapping reveals that the 1-$\sigma$ spread of the scatter in RDs
in the smTFR is $\sim$0.03 dex). In contrast, the total scatter is
found to be significantly larger, with 0.83 dex in the bTFR instead of
0.63 in the smTFR. We used Monte-Carlo simulations to investigate
whether the larger uncertainty associated with the gas mass, compared
to the stellar mass, is responsible for the increase in scatter in the
bTFR. To do this, we simulated 1000 smTFRs by moving the galaxies
within the errorbars in their baryonic mass. We then refitted the
resulting smTFRs, and found that the total scatter (as well as the
scatter for RDs) does not increase significantly. Therefore, it
appears that the bTFR has a significantly larger scatter than the
smTFR, at least for the total galaxy population.

In assessing whether there is an evolution in the intercept of the
bTFR, it is important to account for possible systematic effects,
which are usually more significant than random effects in the TFR (see
P08 and Sect. 3). In the following, we express all systematic
uncertainties in terms of their influence on the evolution of the
zeropoint of the bTFR between z$\sim$0.6 and z=0. Systematic effects
on the velocity estimates in the distant sample were found to be
$\pm$0.02 dex, which corresponds to $\pm$0.08 dex once converted into
$M_{baryonic}$. Regarding the stellar mass, we identified in P08 two
possible systematic effects, which are an evolution in the IMF (+0.05
dex), and an evolution in the K-band mass-to-light ratio (+0.1 dex)
with redshift (see Appendix of P08). Finally, concerning gas masses,
the only possible systematic effect could be an evolution of the SK
law with redshift, since we argued in Sect. 2.3 that this quantity is
derived independently of the IMF. As stated in Sect. 2.3, there is no
evidence of an evolution in the SK law between z$\sim$0.6 and z=0,
based on a comparison between their star-formation densities and those
of local starbursts. Therefore, the main systematic effect that could
affect the baryonic mass would be that related to stellar mass, which
translates into a possible impact on the evolution of the bTFR
zeropoint shift between z$\sim$0.6 and z=0 by +0.15 dex (0.05+0.1, see
above). In conclusion, we find a bTFR zeropoint shift between
z$\sim$0.6 and z=0 of -0.05$\pm$0.08 (random, see above)
$^{+0.23}_{-0.08}$ (systematic).

Using the permutation test of \cite{koen09}, we find a probability of
$\sim$62\% that the local and distant relations have the same slope
and intercept. Given that there are only six RDs in the distant bTFR,
we also tested whether the local and distant intercepts are different,
assuming that the slope does not evolve. To do this, we used a Welch
t-test \citep{koen09}, and found that the probability that they are
equal is 91\%. We also found that this probability drops below 10\%
when we systematically increase the local bTFR zeropoint by at least
0.77 dex: this means that the local and distant bTFR zeropoints are
not statistically different within systematic uncertainties (see
above), assuming no evolution in slope. As stated above, the residual
scatters of the local and distant relation are similar, which suggests
that assuming there is no evolution in slope is a reasonable
hypothesis. Therefore, we conclude that we do not detect any
significant shift in the bTFR between z$\sim$0.6 and z=0.

\section{Discussion \& conclusions}
There is now little doubt that the large scatter found in the
z$\sim$0.6 TFR is caused by non-relaxed systems associated with major
mergers (\citealt{flores06,kassin07,puech08,covington09}; see also
\citealt{kannappan04}). Mergers provide a natural and coherent frame
for interpreting the morpho-kinematics properties of z$\sim$0.6
galaxies (see \citealt{hammer09b}). In this paper, we have reported
that ionized gas extends farther out than the ionizing light from OB
stars: we indeed found a probability of 6\% that the median ratio of
the gas to stellar radius is equal to one. This can be seen as a
natural consequence of shocks produced during these events. It is even
difficult to figure out how gas lying significantly farther out than
any radiation source could be ionized by any other process, given that
a very negligible fraction of objects in the sample are found to
contain evidence of gas outflows, which might spread ionized gas out
to the UV radius.

Returning to the TFR, we confirm that, once restricted to rotating
disks, the z$\sim$0.6 smTFR appears to be shifted toward fainter
masses by 0.34 dex. This shift is found to be very significant within
the random \emph{and} systematic uncertainties. In contrast, by
including the gas fraction of the baryonic mass in the TFR, we found
that, within the uncertainties, the z$\sim$0.6 relation is consistent
with the local relation derived for the \cite{mcgaugh05} sample. This
implies that star formation in rotating disks is mostly fed by gas
that is already gravitationally bound to galaxies, otherwise we would
have detected significant evolution in the bTFR. This does not mean
that there is no external gas accretion: given the still relatively
large associated uncertainties, there could be room for external gas
accretion, at a level of up to roughly one third of the local baryonic
mass. However, there is presently no need for this external gas
accretion.

At higher redshifts, \cite{cresci09} found an evolution of 0.41 dex in
the zeropoint of the smTFR between z$\sim$2.2 and z=0, i.e., of
similar amplitude (within uncertainties) to the evolution found in
this paper between z$\sim$0.6 and z=0. This would imply that the smTFR
does not evolve significantly between z$\sim$2.2 and z$\sim$0.6, while
it evolves significantly between z$\sim$0.6 and z=0. To try to
understand this surprising result, we refitted the smTFR found by
\cite{cresci09}, but assumed that the local slope of reference is that
used in the present study. The choice of the slope in the local
relation is crucial for deriving the evolution in zeropoint as a
function of redshift (see Sect. 5.1 of P08 for details). While the
distant sample used by \cite{cresci09} seems to be representative of
z$\sim$2 galaxies, at least for galaxies with stellar masses higher
than $\sim$2$\times$10$^{10}$M$_\odot$ \citep{forster09}, they used
the local relation derived by \cite{bell01}, which relies on the
sample of \cite{verheijen01} that was shown to be biased toward an
excess of low-mass, gas-rich galaxies by \cite{hammer07}. In P08, we
derived a local smTFR using a representative subsample of the SDSS
from \cite{pizagno07}. In this sample, the slope is found to be
smaller, i.e., 2.8 (see Appendix of P08) instead of the value of 4.5
used by \cite{cresci09}. By refitting the relation of \cite{cresci09}
using the slope derived from this representative local sample, we find
an evolution in zeropoint of $\sim$0.6 dex, instead of $\sim$0.4 dex
found using the \cite{bell01} slope. We conclude that the shift found
by \cite{cresci09} probably underestimates the evolution in zeropoint
between z$\sim$2.2 and z=0.

Finally, \cite{cresci09} interpreted the shift of the z$\sim$2.2 smTFR
as the result of gas accretion onto the forming disks in filaments and
cooling flows, as suggested by theoretical and numerical models. This
gas accretion process could also play a role in feeding the outer
regions of z$\sim$0.6 galaxies with fresh gas, to within a limit of
30\% of the local baryonic mass, as discussed above. However, it would
remain unclear what the ionization source of this gas is (see Sect.
4). These models also clearly predict that these cold flows are
strongly suppressed in the massive, z$\sim$0.6 haloes that the
galaxies studied in this paper inhabit (e.g., \citealt{dekel09}). In a
companion paper \citep{hammer09b}, we explore another possibility that
a large fraction of local spirals could have rebuilt their disk
following a major merger at z$\leq$1. This scenario, dubbed as
``spiral rebuilding disk scenario'', was proposed by \cite{hammer05},
owing to the remarkable coincidence of the evolution of the merger
rate, morphology, and fraction of actively star-forming galaxies. In
this scenario, major mergers expel gas in, e.g., tidal tails, which is
later re-accreted to rebuild a new disk
\citep{barnes02,robertson06,hopkins08}. It is tempting to associate a
significant part of the gas reservoir with this process, which can
therefore provide us with an evolutionary framework in which to
intepret the non-evolution of the bTFR with redshift, and the finding
that ionized gas extends farther out than the stellar UV light. A
dynamical imprint of this accretion might already have been detected
in terms of the lower V/$\sigma$ found in distant gaseous disks
compared to local spirals \citep{puech07}. Therefore, different
physical mechanisms could be driving galaxy evolution and the shift in
the smTFR, depending on redshift, which would complicate the
interpretation of the evolution in the smTFR from very high redshift
to z=0.

\begin{acknowledgements}
This study made use of GOODS data in the CDFS field (MIPS and HST/ACS
images), as well as of the 2MASS and the CDS databases. We thank X.
Hernandez and L. Chemin for useful discussions about the subject of
this paper.
\end{acknowledgements}


\begin{thebibliography}{}
  \bibitem[Atkinson et al.(2007)]{atkinson07} Atkinson, N., Conselice,
    C.J., \& Fox, N. 2007, \mnras, submitted, astro-ph/0712.1316
  \bibitem[Avila-Reese et al.(2008)]{avila08} Avila-Reese, V., Zavala,
    J., Firmani, C., et al. 2008, \aj, 136, 1340
  \bibitem[Bamford et al.(2007)]{bamford07} Bamford, S.P.,
    Milvang-Jensen, B., \& Aragon-Salamanca, A. 2007, \mnras, 378, L6
  \bibitem[Barnes(2002)]{barnes02} Barnes, J.E. 2002, \mnras, 333, 481
  \bibitem[Begum et al.(2008)]{begum08} Begum, A., Chengalur, J.N.,
    Karachentsev, I.D., et al. 2008, \mnras, 386, 138
  \bibitem[Bell \& de Jong(2001)]{bell01} Bell, E.F., \& de Jong, R.S.
    2001, \apj, 550, 212
  \bibitem[Bell et al.(2003)]{bell03} Bell, E.~F., McIntosh, D.~H.,
  Katz, N., \& Weinberg, M.~D.\ 2003, \apjs, 149, 289
\bibitem[Bertin \& Arnouts(1996)]{bertin96} Bertin, E., \& Arnouts,
  S.\ 1996, \aaps, 117, 393
  \bibitem[Borch et al.(2006)]{borch06} Borch, A., Meisenheimer, K.,
  Bell, E.F., et al. 2006, \aap, 453, 869
  \bibitem[Bouch\'e et al.(2007)]{bouche07} Bouché, N., Cresci, G.,
    Davies, R., et al. 2007, \apj, 671, 303
  \bibitem[Bruzual \& Charlot(2003)]{bruzual03} Bruzual, G., \&
    Charlot, S.\ 2003, \mnras, 344, 1000
  \bibitem[Chary \& Elbaz(2001)]{chary01} Chary, R., Elbaz, D.\ 2001,
  \apj, 556, 562
  \bibitem[Conselice et al.(2005)]{conselice05} Conselice, C.~J.,
Bundy, K., Ellis, R.~S., Brichmann, J., Vogt, N.~P., \& Phillips,
A.~C.\ 2005, \apj, 628, 160
  \bibitem[Covington et al.(2009)]{covington09} Covington, M.D.,
    Kassin, S.A., Dutton, A.A., et al. 2009, \apj, submitted,
    astro-ph/0902.0566
  \bibitem[Cresci et al.(2009)]{cresci09} Cresci, G., Hicks, E.K.S.,
    Genzel, R., et al. 2009, \apj, 697, 115
  \bibitem[De Rijcke et al.(2007)]{derijcke07} De Rijcke, S.,
    Zeilinger, W.~W., Hau, G.~K.~T., Prugniel, P., \& Dejonghe,
    H.\ 2007, \apj, 659, 1172
  \bibitem[Dekel et al.(2009)]{dekel09} Dekel, A., \& Birnboim, Y.,
    Engel, G., et al. 2009, Nature, 457, 451
  \bibitem[Erb et al.(2006)]{erb06} Erb, D.K., Steidel, C.C., Shapley,
    A.E., et al. 2006, \apj, 646, 107
  \bibitem[Flores et al.(2006)]{flores06} Flores, H., Hammer, F.,
    Puech, M., Amram, P., \& Balkowski, C.\ 2006, \aap, 455, 107
  \bibitem[F{\"o}rster Schreiber et al.(2009)]{forster09} Forster
    Schreiber, N.~M., et al.\ 2009,\apj, accepted, arXiv:0903.1872
  \bibitem[Geha et al.(2006)]{geha06} Geha, M., Blanton, M.~R.,
    Masjedi, M., \& West, A.~A.\ 2006, \apj, 653, 240
  \bibitem[Goruvich et al.(2004)]{gurovich04} Gurovich, S., McGaugh,
    S.S., Freeman, K.C., et al. 2004, PASA, 21, 412
  \bibitem[Hammer et al.(2005)]{hammer05} Hammer, F., Flores, H.,
  Elbaz, D., Zheng, X.~Z., Liang, Y.~C., \& Cesarsky, C.\ 2005, \aap,
  430, 115
  \bibitem[Hammer et al.(2007)]{hammer07} Hammer, F., Puech, M.,
    Chemin, L., Flores, H., \& Lehnert, M.\ 2007, \apj, 662, 322
  \bibitem[Hammer et al.(2009)]{hammer09} Hammer, F., Flores, H.,
    Yang, Y., et al. 2009, \aap, 496, 381
  \bibitem[Hammer et al.(2009b)]{hammer09b} Hammer, F., et al. 2009,
    A\&A, accepted, astro-ph/0903.3962
  \bibitem[Hopkins et al.(2009)]{hopkins08} Hopkins, P.F., Cox, T.J.,
    Younger, J.D., \& Hernquist, L. 2009, \apj, 691, 1168
  \bibitem[Kannappan \& Barton(2004)]{kannappan04} Kannappan, S.~J.,
\& Barton, E.~J.\ 2004, \aj, 127, 2694
  \bibitem[Kassin et al.(2007)]{kassin07} Kassin, S.A., Weiner, B.J.,
  Faber, S.M., et al. 2007, \apj, 660, 35
  \bibitem[Kennicutt(1989)]{kennicutt89} Kennicutt, R.\ 1989, \apj,
  344, 685
  \bibitem[Kennicutt(1998)]{kennicutt98} Kennicutt, R.\ 1998, \araa,
  36, 189
  \bibitem[Koen \& Lombard(2009)]{koen09} Koen, C., \& Lombard, F.
    2009, \mnras, 395, 1657
  \bibitem[Lah et al.(2007)]{lah07} Lah, P., Chengalur, J.N., Briggs,
    F.H., et al. 2007, \mnras, 376, 1357
  \bibitem[Le Floc'h et al.(2005)]{lefloch05} Le Floc'h, E., Papovich,
  C., Dole, H. et al. 2005, \apj, 632, 169
  \bibitem[Mannucci et al.(2009)]{mannucci09} Mannucci, F., Cresci,
    G., Maiolino, R., et al. 2009, \mnras, submitted,
    astro-ph/0902.2398
  \bibitem[Maraston et al.(2006)]{maraston06} Maraston, C., Daddi, E.,
    Renzini, A., et al. 2006, \apj, 652, 85
  \bibitem[Marchesini et al.(2009)]{marchesini09} Marchesini, D., van
    Dokkum, P.~G., F{\"o}rster Schreiber, N.~M., Franx, M., Labb{\'e},
    I., \& Wuyts, S.\ 2009, \apj, 701, 1765
  \bibitem[McGaugh(2000)]{mcgaugh00} McGaugh, S.S. 2000, \apj, 533,
    L99
  \bibitem[McGaugh(2004)]{mcgaugh04} McGaugh, S.~S.\ 2004, \apj, 609,
    652
  \bibitem[McGaugh(2005)]{mcgaugh05} McGaugh, S.S. 2005, \apj, 632,
    859
  \bibitem[Meyer et al.(2008)]{meyer08} Meyer, M.J., Zwaan, M.A.,
    Webster, R.L. 2008, \mnras, 391, 1712
  \bibitem[Montero-Ibero et al.(2006)]{montero06} Montero-Ibero, A.,
    Arribas, S., Colina, L. 2006, \apj, 637, 138
  \bibitem[Neichel et al.(2008)]{neichel08} Neichel, B., et al. 2008,
  \aap, 484, 159
  \bibitem[Noordermeer\&Verheijen(2007)]{noordermeer07} Noordermeer,
    E., \& Verheijen, M.A.W. 2007, \mnras, 381, 1463
  \bibitem[Peirani et al.(2009)]{peirani09} Peirani, S., Hammer, F.,
    Flores, H., et al. 2009, \aap, 496, 51
  \bibitem[Persic \& Salucci(1991)]{persic91} Persic M. \& Salucci P.
  1991, \apj, 368, 60
  \bibitem[Pizagno et al.(2007)]{pizagno07} Pizagno, J., Prada, F.,
  Weinberg, D.H., et al. 2007, \aj, 134, 945
  \bibitem[Puech et al.(2006)]{puech06} Puech, M., Hammer, F., Flores,
  H., {\"O}stlin, G., \& Marquart, T.\ 2006b, \aap, 455, 119
  \bibitem[Puech et al.(2007)]{puech07} Puech, M., Hammer, F.,
  Lehnert, M.~D., \& Flores, H.\ 2007a, \aap, 466, 83
  \bibitem[Puech et al.(2007b)]{puech07b} Puech, M., Hammer, F.,
    Flores, H., et al. 2007b, \aap, 476, 21
  \bibitem[Puech et al.(2008)]{puech08} Puech, M., Flores, H., Hammer,
  F., et al. 2008, \aap, 484, 173
  \bibitem[Puech et al.(2009)]{puech09} Puech, M., Hammer, F., Flores,
    H., et al. 2009, \aap, 493, 899
  \bibitem[Pozzetti et al.(2007)]{pozzetti07} Pozzetti, L.,
    Bolzonella, M., Lamareille, F., et al. 2007, \aap, 474, 443
  \bibitem[Ravikumar et al.(2007)]{ravi07} Ravikumar, C.~D., et al.\
  2007, \aap, 465, 1099
  \bibitem[Robertson et al.(2006)]{robertson06} Robertson, B.,
  Bullock, J.~S., Cox, T.~J., Di Matteo, T., Hernquist, L., Springel,
  V., \& Yoshida, N.\ 2006, \apj, 645, 986
  \bibitem[Rodrigues et al.(2008)]{rodrigues08} Rodrigues, M., Hammer,
    F., Flores, H., et al. 2008, \aap, 492, 371
  \bibitem[Schiminovich(2008)]{schim08} Schiminovich, D.\ 2008, The
    Evolution of Galaxies Through the Neutral Hydrogen Window, 1035,
    180
  \bibitem[Stark et al.(2009)]{stark09} Stark, D.~V., McGaugh, S.~S., \&
  Swaters, R.~A.\ 2009, \aj, 138, 392
  \bibitem[Str\"omberg(1939)]{stromberg39} Strömberg, B. 1939, \apj,
    89, 526
  \bibitem[Trachternach et al.(2009)]{trachternach09} Trachternach,
    C., de Blok, W.~J.~G., McGaugh, S.~S., van der Hulst, J.~M., \&
    Dettmar, R.~-.\ 2009, \aap, 505, 577
  \bibitem[Verheijen(2001)]{verheijen01} Verheijen, M.A.W. 2001, \apj,
    563, 694
  \bibitem[Weiner et al.(2006)]{weiner06} Weiner, B.~J., et al.\ 2006,
  \apj, 653, 1027
  \bibitem[Yang et al.(2008)]{yang07} Yang, Y., Flores, H., Hammer,
    F., et al. 2008, \aap, 477, 789


\end{thebibliography}
\end{document}